\documentclass[iop]{emulateapj}

\newcommand{\eqref}[1]{(\ref{#1})}

\def\lesssim{\mathrel{\hbox{\rlap{\hbox{\lower4pt\hbox{$\sim$}}}\hbox{$<$}}}}
\def\gtrsim{\mathrel{\hbox{\rlap{\hbox{\lower4pt\hbox{$\sim$}}}\hbox{$>$}}}}

\shorttitle{LkHa 330}
\shortauthors{Akiyama et al.}

\usepackage{color}
\usepackage{ulem}
\begin{document}


\title{SPIRAL STRUCTURE AND DIFFERENTIAL DUST SIZE DISTRIBUTION IN THE LkH$\alpha$ 330 DISK}


\author{EIJI AKIYAMA\altaffilmark{1}, JUN HASHIMOTO\altaffilmark{1}, HAUYU BAOBABU LIU\altaffilmark{2,4}, JENNIFER I-HSIU LI\altaffilmark{3,4}, MICHAEL BONNEFOY\altaffilmark{5,6}, RUOBING DONG\altaffilmark{7}, YASUHIRO HASEGAWA\altaffilmark{1,8}, THOMAS HENNING\altaffilmark{6},  MICHAEL L. SITKO\altaffilmark{9,10}, MARKUS JANSON\altaffilmark{11}, MARKUS FELDT\altaffilmark{6}, JOHN WISNIEWSKI\altaffilmark{12}, TOMOYUKI KUDO\altaffilmark{13}, NOBUHIKO KUSAKABE\altaffilmark{14}, TAKASHI TSUKAGOSHI\altaffilmark{15}, MUNETAKE MOMOSE\altaffilmark{15}, TAKAYUKI MUTO\altaffilmark{16}, TETSUO TAKI\altaffilmark{17}, MASAYUKI KUZUHARA\altaffilmark{14}, SATOSHI MAYAMA\altaffilmark{18}, MICHIHIRO TAKAMI\altaffilmark{4}, NAGAYOSHI OHASHI\altaffilmark{13}, CAROL A. GRADY\altaffilmark{19,20}, JUNGMI KWON\altaffilmark{21}, CHRISTIAN THALMANN\altaffilmark{22}, LYU ABE\altaffilmark{23}, WOLFGANG BRANDNER\altaffilmark{6}, TIMOTHY D. BRANDT\altaffilmark{24}, JOSEPH C. CARSON\altaffilmark{6,25}, SEBASTIAN EGNER\altaffilmark{13}, MIWA GOTO\altaffilmark{26}, OLIVIER GUYON\altaffilmark{13}, YUTAKA HAYANO\altaffilmark{13}, MASAHIKO HAYASHI\altaffilmark{1,18}, SAEKO S. HAYASHI\altaffilmark{13}, KLAUS W. HODAPP\altaffilmark{27}, MIKI ISHII\altaffilmark{1}, MASANORI IYE\altaffilmark{1}, GILLIAN R. KNAPP\altaffilmark{24}, RYO KANDORI\altaffilmark{1}, TARO MATSUO\altaffilmark{28}, MICHAEL W. MCELWAIN\altaffilmark{29}, SHOKEN MIYAMA\altaffilmark{30}, JUN-ICHI MORINO\altaffilmark{1}, AMAYA MORO-MARTIN\altaffilmark{31,32}, TETSUO NISHIMURA\altaffilmark{13}, TAE-SOO PYO\altaffilmark{13}, EUGENE SERABYN\altaffilmark{8}, TAKUYA SUENAGA\altaffilmark{1,33}, HIROSHI SUTO\altaffilmark{1}, RYUJI SUZUKI\altaffilmark{1}, YASUHIRO H. TAKAHASHI\altaffilmark{1,21}, NARUHISA TAKATO\altaffilmark{13}, HIROSHI TERADA\altaffilmark{13}, DAIGO TOMONO\altaffilmark{13}, EDWIN L. TURNER\altaffilmark{24,34}, MAKOTO WATANABE\altaffilmark{35}, TORU YAMADA\altaffilmark{36}, HIDEKI TAKAMI\altaffilmark{1}, TOMONORI USUDA\altaffilmark{1}, MOTOHIDE TAMURA\altaffilmark{1,14,21}}

\affil{
\altaffilmark{1}National Astronomical Observatory of Japan, 2-21-1, Osawa, Mitaka, Tokyo, 181-8588, Japan; eiji.akiyama@nao.ac.jp \\
\altaffilmark{2}European Southern Observatory, Karl Schwarzschild Str 2, 85748 Garching bei M\"{u}nchen, Germany \\
\altaffilmark{3}Department of Astronomy, University of Illinois at Urbana-Champaign 1002 West Green Street, Urbana, IL 61801, USA \\
\altaffilmark{4}Institute of Astronomy and Astrophysics, Academia Sinica, P.O. Box 23-141, Taipei, 10617, Taiwan \\
\altaffilmark{5}Univ. Grenoble Alpes, IPAG, F-38000 Grenoble, France. CNRS, IPAG, F-38000 Grenoble, France \\
\altaffilmark{6}Max Planck Institute for Astronomy, K\"{o}nigstuhl 17, 69117, Heidelberg, Germany \\
\altaffilmark{7}Lawrence Berkeley National Laboratory, Berkeley, CA 94720, USA \\
\altaffilmark{8}Jet Propulsion Laboratory, California Institute of Technology, 4800, Oak Grove Drive, Pasadena, CA, 91109, USA \\
\altaffilmark{9}Space Science Institute, 4750 Walnut St., Suite 205, Boulder, CO 80301, USA \\
\altaffilmark{10}Department of Physics, University of Cincinnati, Cincinnati, OH 45221-0011, USA \\
\altaffilmark{11}Department of Astronomy, Stockholm University, AlbaNova University Center, Stockholm, 106 91, Sweden \\
\altaffilmark{12}Department of Physics and Astronomy, The University of Oklahoma, 440 W. Brooks St. Norman, OK  73019, USA \\
\altaffilmark{13}Subaru Telescope, National Astronomical Observatory of Japan, 650, North A'ohoku Place, Hilo, HI  96720, USA \\
\altaffilmark{14}Astrobiology Center of NINS, 2-21-1, Osawa, Mitaka, Tokyo, 181-8588, Japan \\
\altaffilmark{15}College of Science, Ibaraki University, 2-1-1, Bunkyo, Mito, Ibaraki, 310-8512, Japan \\
\altaffilmark{16}Division of Liberal Arts, Kogakuin University, 1-24-2, Nishi-Shinjuku, Shinjuku-ku, Tokyo, 163-8677, Japan \\
\altaffilmark{17}Department of Earth and Planetary Sciences, Tokyo Institute of Technology, 2-12-1, Ookayama, Meguro-ku, Tokyo 152-8551, Japan \\
\altaffilmark{18}The Center for the Promotion of Integrated Sciences, The Graduate University for Advance Studies, Shonan International Village Hayama-cho, Miura-gun, Kanagawa, 240-0115, Japan \\
\altaffilmark{19}Exoplanets and Stellar Astrophysics Laboratory, Code 667, Goddard Space Flight Center, Greenbelt, MD 20771, USA \\
\altaffilmark{20}Eureka Scientific, 2452 Delmer, Suite 100, Oakland, CA 96002, USA \\
\altaffilmark{21}Department of Astronomy, The University of Tokyo, 7-3-1, Hongo, Bunkyo-ku, Tokyo, 113-0033, Japan \\
\altaffilmark{22}Institute for Astronomy, ETH Zurich, Wolfgang-Pauli-Strasse 27, 8093, Zurich, Switzerland \\
\altaffilmark{23}Laboratoire Hippolyte Fizeau, UMR6525, Universite de Nice Sophia-Antipolis, 28, avenue Valrose, 06108, Nice Cedex 02, France \\
\altaffilmark{24}Department of Astrophysical Sciences, Princeton University, Peyton Hall, Ivy Lane, Princeton, NJ 08544, USA \\
\altaffilmark{25}Department of Physics and Astronomy, College of Charleston, 66 George St., Charleston, SC 29424, USA \\
\altaffilmark{26}12 Universit$\ddot{\rm a}$ts-Sternwarte M$\ddot{\rm u}$nchen, Ludwig-Maximilians-Universit $\ddot{\rm a}$t, Scheinerstr. 1, 81679 M$\ddot{\rm u}$nchen, Germany \\
\altaffilmark{27}Institute for Astronomy, University of Hawaii, 640 North A'ohoku Place, Hilo, HI, 96720, USA \\
\altaffilmark{28}Department of Astronomy, Kyoto University, Kita-shirakawa-Oiwake-cho, Sakyo-ku, Kyoto, 606-8502, Japan \\
\altaffilmark{29}Exoplanets and Stellar Astrophysics Laboratory, Code 667, Goddard Space Flight Center, Greenbelt, MD, 20771, USA \\
\altaffilmark{30}Hiroshima University, 1-3-2, Kagamiyama, Higashi-Hiroshima, Hiroshima, 739-8511, Japan \\
\altaffilmark{31}Space Telescope Science Institute, 3700 San Martin Drive, Baltimore, MD, 21218, USA \\
\altaffilmark{32}Center for Astrophysical Sciences, Johns Hopkins University, Baltimore, MD, 21218, USA \\
\altaffilmark{33}Department of Astronomical Science, School of Physical Sciences, Graduate University for Advanced Studies (SOKENDAI), Mitaka, Tokyo, 181-8588, Japan \\
\altaffilmark{34}Kavli Institute for the Physics and Mathematics of the Universe, The University of Tokyo, 5-1-1, Kashiwanoha, Kashiwa, Chiba, 227-8568, Japan \\
\altaffilmark{35}Department of Cosmosciences, Hokkaido University, Kita-ku, Sapporo, 060-0810, Japan \\
\altaffilmark{36}Astronomical Institute, Tohoku University, Aoba-ku, Sendai, Miyagi, 980-8578, Japan 
}

\begin{abstract}
Dust trapping accelerates the coagulation of dust particles, and thus it represents an initial step toward the formation of planetesimals. We report $H$-band (1.6 $\mu$m) linear polarimetric observations and 0.87 mm interferometric continuum observations toward a transitional disk around LkH$\alpha$ 330. As results, a pair of spiral arms were detected in the $H$-band emission and an asymmetric (potentially arm-like) structure was detected in the 0.87 mm continuum emission. We discuss the origin of the spiral arm and the asymmetric structure, and suggest that a massive unseen planet is the most plausible explanation. The possibility of dust trapping and grain growth causing the asymmetric structure was also investigated through the opacity index ($\beta$) by plotting the observed SED slope between 0.87 mm from our SMA observation and 1.3 mm from literature. The results imply that grains are indistinguishable from ISM-like dust in the east side ($\beta = 2.0\pm0.5$), but much smaller in the west side $\beta = 0.7^{+0.5}_{-0.4}$, indicating differential dust size distribution between the two sides of the disk. Combining the results of near-infrared and submillimeter observations, we conjecture that the spiral arms exist at the upper surface and an asymmetric structure resides in the disk interior. Future observations at centimeter wavelengths and differential polarization imaging in other bands (Y to K) with extreme AO imagers are required to understand how large dust grains form and to further explore the dust distribution in the disk.

\end{abstract}




\section{Introduction}
Dust grains are the raw material for planetesimals, which over time form rocky planet cores. The widely accepted core accretion scenario implies that coagulation of micron-sized dust grains into larger bodies is a first step for planet formation. Thus, it is important to investigate the grain size and its distribution in the birthplaces of planets: protoplanetary disks. If the emission is optically thin, the opacity index ($\beta$) can be used as an observational indication of grain size (possibly grain growth), where $\kappa_d$ $(\propto \lambda^{-\beta})$ is the dust opacity and $\lambda$ is the wavelength. A number of observations reveal that the value of $\beta$ is generally low for circumstellar disks around low and intermediate-mass young stellar objects (YSOs) \citep{beckwith91,dalessio01,andrews05,draine06,ricci10a,ricci10b,guilloteau11,menu14}.
The exploration of radial and azimuthal dust distribution in various sizes of dust is essential for revealing where and how dust grain coagulation takes place. Since the $\beta$ offers suggestive evidence for the amount of dust grains whose size is approximately comparable to the wavelength of observation, the analysis of $\beta$ and its distribution help understand the initial condition for the subsequent formation of planetesimals. 
 
We have conducted $H$-band linear polarimetric observations that trace (sub)micron-sized dust at the disk scattering surface and 0.87 mm dust continuum observations that detect the thermal emission from approximately millimeter-sized grains around the disk mid-plane. LkH$\alpha$ 330 is a G3 star \citep{cohen79} of 2.5 M$_\sun$ \citep{salyk09} and 15 L$_\sun$ \citep{andrews11} located in the Perseus molecular clouds, which is about 250 pc away from the Sun \citep{enoch06}. The disk of LkH$\alpha$ 330 is one of the 4 transitional disks found in the Spitzer Space Telescope Cores to Disks (c2d) Legacy Program, where they took $5-35$ $\mu$m spectra with the Infrared Spectrograph (IRS) for 100 stellar systems \citep{brown07}. Later, \citet{brown08,brown09} constructed the first resolved submillimeter image of the disk with an angular resolution of 0\farcs3 using the Submillimeter Array (SMA) at 340 GHz. This observation directly confirmed the cavity in the dust disk. Then, a two-dimensional Monte Carlo radiative transfer code \citep[RADMC;][]{dullemond04} was used to model the disk. In this model, the disk is flared, density is a power law of radius, and the cavity is represented by the inner and outer gap radius and a density reduction factor. In addition to the best-fitting model of the spectral energy distribution (SED),  they indicated that the radius of the cavity was 47 AU and that hot gas resided within 0.8 AU. \citet{andrews11} re-calibrated the data from \citet{brown08,brown09} and combined it with SMA compact (COM) data. They treated the surface density by the similarity solution with an exponential tail in their analysis and derived the cavity radius to be 68 AU. 

\citet{isella13} also studied the asymmetric component in the LkH$\alpha$ 330 disk with SMA archival data and CARMA data at $\lambda = 1.3$ mm. As results of the visibility fitting with parametric gaussian disk models, they found that the asymmetric component traces a narrow circular arc, which extends in the azimuthal direction by about 90$\arcdeg$ and accounts for about 1/3 of the total disk flux at $\lambda = 1.3$ mm. They discuss possible mechanisms of the arc formation including planets, Rossby waves instabilities (RWI), baroclinic instabilities, disk warping, and disk shadowing. Observations of molecular tracers might help to further confirm the origin of the cavity in the disk.

In this paper, the details of the $H$-band linear polarimetric observations and 0.87 mm dust continuum observations of LkH$\alpha$ 330 are provided in section \ref{obs}, and the obtained images which show a spiral arm and asymmetric structures in the LkH$\alpha$ 330 disk are presented in section \ref{result}. In section \ref{discussion}, micron and millimeter-sized dust distributions and the origin of the spiral arm as well as the asymmetric structure are discussed. Finally, we examine the possibility of dust trapping by estimating the dust opacity index using our observations and the data from the literature. Section \ref{summary} summarizes our findings.     

\section{Observations and Data Reduction}
\label{obs}
\subsection{Subaru/HiCIAO $H$-band Observation}
\label{subaru}
$H$-band (1.6 $\mu$m) linear polarimetric observations of LkH$\alpha$ 330 were conducted on December 22nd 2011 (UT) as a part of the Strategic Explorations of Exoplanets and Disks with Subaru \citep[SEEDS;][]{tamura09} project. The High Contrast Instrument for the Subaru Next Generation Adaptive Optics \citep[HiCIAO;][]{tamura06} with a dual-beam polarimeter on the Subaru 8.2 m telescope was used. The polarimetric observation mode acquires the \textit{o}-ray and \textit{e}-ray simultaneously, and images a 10\arcsec $\times$ 20\arcsec field of view (FOV) with a pixel scale of 9.50 mas/pixel. The adaptive optics system \citep[AO188;][]{hayano08} functioned mostly stable during observation, providing stellar point-spread function (PSF) with a FWHM of 0\farcs15 in the \textit{H}-band. The adaptive optics corrections functioned well in the observation and achieve a saturated radius (inner working angle) of 0\farcs2, corresponding to $\sim$ 50 AU. As discussed in Section \ref{result}, the region less than 0\farcs2 from the central star in principle shows no saturation problems nor discernible misalignment of polarization vectors and thus could potentially be usable. However, we did not include the region within $r = 0\farcs2$, and to be conservative, polarization measurements at radii more distant than 0\farcs2 from the central star are regarded as more accurate and safe analysis. The data sets were taken with a combination of Angular Differential Imaging (ADI) and Polarimetric Differential Imaging (PDI) mode with a net field rotation of 12 degrees. A double Wollaston prism system was used for splitting the incident light, and as a result, four images with individual FOV of 5\arcsec $\times$ 5\arcsec are generated. The total integration time is 1600 s for the final PI image. All reduced data were corrected for field rotation and then integrated. Note that only data obtained by PDI mode were used and ADI analysis leaves for our future work.

A standard reduction method \citep{hinkley09} was applied to the polarimetric data, including corrections for distortion and flat-field, using the Image Reduction and Analysis Facility (IRAF\footnote{IRAF is distributed by National Optical Astronomy Observatory, which is operated by the Association of Universities for Research in Astronomy, Inc., under cooperative agreement with the National Science Foundation.}) software as in previous SEEDS publications \citep[e.g.][]{hashimoto11}. Calibration for instrumental polarization of the Nasmyth HiCIAO instrument was made based on the Mueller-matrices technique, which quantizes polarization characteristics \citep{joos08}. Finally, a PI image was calculated from $\sqrt{Q^2+U^2}$, where $Q$ and $U$ are orthogonal linear polarizations.

\subsection{Submillimeter Array Observation}
\label{sma}
LkH$\alpha$ 330 was observed at 343.796 GHz, corresponding to 0.87 mm in wavelength, by the SMA on December 18th 2014 (UT). The very extended configuration was selected, with eight 6 m diameter antennae whose baseline lengths ranged between 30 m and 590 m. The double sideband (DSB) receivers with a total bandwidth of 4 GHz was used  to maximize the sensitivity to the continuum. The bright quasar 3c279 was used to calibrate the bandpass and Callisto was observed for determining an absolute flux scale, whose uncertainty is estimated to be approximately 10$\%$. The nearby quasar 3c84 was also observed to calibrate the atmospheric and instrumental phase before and after observing the target source. The system temperature was between 400 K and 600 K and the precipitable water vapor (PWV) was less than 2.5 mm during the observation. The integration time on source was 7.4 hours. All of the calibrations were done using the MIR software package\footnote{The package is open for public use and can be obtained from https://www.cfa.harvard.edu/~cqi/mircook.html.}. The deconvolution was done using the MIRIAD software and the natural weighting was applied when we CLEANed the image so that we can obtain higher sensitivity. Self-calibration was also introduced to the imaging by applying the continuum image itself as a model.

\section{Results}
\label{result}
The polarized intensity image in $H$-band of the transitional disk around LkH$\alpha$ 330 is presented in Figure \ref{fig:fig1}a. The figure shows two bright components at approximately 70 AU in the east and west side of the disk, whose intensities are 6.1 and 3.6 mJy arcsec$^{-2}$, respectively. The rms is only 0.16 mJy arcsec$^{-2}$, producing S/N ratios of 38 and 22, respectively. The western bright component includes a spiral arm structure extending to the southeast direction, and the bright component in the eastern side also seems to include a spiral arm-like structure, although its significance is less clear than the other side. 

Figure \ref{fig:fig1}b displays the continuum image at  $\lambda$ = 0.87 mm obtained from the SMA. The synthesized beam of $0\farcs31 \times 0\farcs27$ was obtained as shown in the lower left corner in the same Figure. The rms in this image is 1.67 mJy beam$^{-1}$. The emission extends 0\farcs6 in radius, corresponding to 150 AU, and a surface brightness depletion is seen around the central star. As in the $H$-band polarized image shown in Figure \ref{fig:fig1}a, asymmetric structure can be seen at the region of $r \la 1\arcsec$ from the central star. The continuum peak in the west side is 29.6 $\pm$ 1.6 mJy beam$^{-1}$, while the one in the east side is 24.8 $\pm$ 1.6 mJy beam$^{-1}$. The continuum image superimposed over the $H$-band polarized image in a larger view is provided in Figure \ref{fig:fig1}c. Both of the emissions are similarly distributed, and their peak intensities spatially match with each other.

Since the Strehl ratio provided by the AO was modest (approximately 0.2), the seeing PSF was not perfectly corrected for, and thus a stellar halo partially remains in the image. The polarization by forward and backward scattering in the disk becomes smaller because the polarization direction significantly deviates from 90$\arcdeg$ of the scattering angle, making the halo appear more polarized along the minor axis of the disk. Figure \ref{fig:fig2} displays how the polarized halo affects the PI image. Figure \ref{fig:fig2}a shows the original PI image with no polarized halo correction. The polarization vectors align along the minor axis due to the polarized halo. However, the polarization vectors become circularly aligned when the remaining polarized halo is subtracted from the PI image as shown in Figure \ref{fig:fig2}b. The correction is made by measuring the Stokes parameters, I, Q, and U at the region between 0\farcs2375 and 2\farcs85 (corresponding to 25 to 300 pixels). The uniformly distributed polarized halo, which has a polarization degree of $0.51 \pm 0.01$ and a polarization angle of $172.7 \pm 0.5\degr$, is subtracted from the image in Figure \ref{fig:fig2}a. The correction method is described in detail by  \citet{hashimoto12}. Note that the region beyond $r = 0\farcs2$ can be safely analyzed after the correction, since the PI of the disk is now distinguished from the polarized halo and no misalignment of polarization vectors can be seen beyond $r = 0\farcs2$.

We further checked the effect of instrumental polarization by the radial polarization (${\rm U}_\phi$) image provided in the Appendix. The polarization component that is not originated from the source is approximately 10 times smaller than the component polarized by the disk source. As a result, the effect of the instrumental polarization component is negligible in our results and therefore not affecting the discussed structures significantly. See more details in the Appendix.


\section{Discussion} 
\label{discussion}

\subsection{Dust distribution and implications of a massive planet} 
\label{dist}
Figure \ref{fig:fig3}a shows the radial surface brightness profiles along the directions of P. A. = 94$\arcdeg$ and 235$\arcdeg$ crossing the scattered emission peaks and P. A. = 258$\arcdeg$ crossing the continuum peak in the radial direction as denoted by red lines in Figure \ref{fig:fig1}c. Both of the scattering emission peaks are found at $r \sim 0\farcs25$ ($\sim$ 62.5 AU) from the central star indicated by dashed circle in Figure \ref{fig:fig1}c. Note that the polarized emission peaks at P. A. = 94$\arcdeg$ and 235$\arcdeg$ are almost located along the direction of semi-major axis (the position angle of LkH$\alpha$ 330 is 80$\arcdeg$ \citep{andrews11} measured from the north in counter-clockwise). While surface brightness asymmetries along the semi-minor axis can be strongly affected by anisotropic scattering, asymmetries along the semi-major axis are unaffected by the scattering efficiency and thus reflect intrinsic variations in the disk structure. Thus, the disk inclination of 35$\arcdeg$ \citep{isella13} does not affect the results very much. 
Figure \ref{fig:fig3}b represents the azimuthal surface brightness profile at $r = 0\farcs25$ and shows a non-axisymmetric torus in the disk. The surface brightnesses in the west and east sides are approximately 3 and 6 times higher than the base levels found in the north and south regions. 

As shown in Figure \ref{fig:fig1}c, the two continuum peaks at the east and west sides of the disk match with the peaks seen in the $H$-band PI image, indicating that small-sized dust (0.1 - 1 $\mu$m) has a similar distribution to the millimeter sized dust when phase function including polarization property is the same everywhere. The radial profiles of scattering and continuum emissions also support that they similarly distribute in the disk because their own peaks are seen at the same radial distance of 0$\farcs25$ as shown in Figure \ref{fig:fig3}a. It is important to know whether the disk is optically thick or not at $\lambda = 0.87$ mm continuum when they are compared with other emissions at different wavelength. It can be roughly checked by comparing the brightness temperature ($T_{\rm b}$) with the radiative equilibrium temperature. As a result, $T_{\rm b}$ at the continuum peak was 9.6 K, which is much lower than the radiative equilibrium temperature of 98 K at $r = 0\farcs25$, indicating that the continuum emission is optically thin. Our results suggest that dust grains between (sub)micron and millimeter in size accumulate at roughly the same radius, but not necessarily from an annular structure of equal thickness.

Both the $H$-band PI image and the 0.87 mm continuum image shown in Figure \ref{fig:fig1}c indicate that there is a dust concentration in the disk. Recent theoretical studies have proposed several mechanisms to trap dust particles: snow lines, dead zones generated by magneto-rotational instability \citep[MRI;][]{balbus91,dzyurkevich10}, anticyclonic vortices that generate high pressure region triggered by RWI \citep{meheut12}, and pressure bumps generated by massive planets \citep{{paardekooper04},zhu12,pinilla12a,pinilla12b,pinilla15}. Among such possibilities, LkH$\alpha$ 330 implies the possible presence of a massive planet because at least spiral structure is associated in the disk. Spiral arms have been discovered in other transitional disks, such as SAO 206462, MWC 758, and HD 100453 and theoretical simulations reproduce the observed spiral features by embedded planet(s), indicating that spiral arms can be induced by planet-disk interaction \citep{muto12,garufi13,grady13,benisty15,dong15,dong16,wagner15}. 

A pair of arm-like features in the east and west sides of the LkH$\alpha$ 330 disk, extending from a continuum peak toward the southeast and northwest, may signify the presence of a massive planet probably located in the south east or north west direction as an extension of the spiral arms and which can generate a pressure bump that locally accumulates a large amount of dust. \citet{dong15} produced the $H$-band polarized intensity images by the newly developed Athena $+ +$ MHD code (J. M. Stone et al. 2016, in preparation) showed spiral structure with $m = 2$ mode (see Figure 2 in \citet{dong15}) by a single massive planet at outer disk region. They also predicts that the brightness of the observed arm is expected to be enhanced by 100 to 300 \% from the emission free background, which is consistent with our result of 380 \% enhancement. 

\citet{isella13} showed that the interaction between the disk and planet is a possible mechanism by utilizing hydrodynamic simulation code FARGO \citep{masset00} as a means to investigate the origin of the asymmetric disk structure. They explored disk structure during 100 to 2000 disk rotations with viscosity of 0.002 and 0.02 and the number of planets with different mass between 3 and 10 $M_J$, and could reproduce the observed azimuthal asymmetries at $\lambda$ = 1.3 mm emission. \citet{birnstiel13} also demonstrated the formation of azimuthal asymmetries in the surface density by disk-planet interactions during their orbital rotation. Therefore, large grains are probably trapped or piled up by a pressure bump generated by a massive planet that depletes gas surface density significantly while interacting with the gas disk \citep{pinilla12a, pinilla12b}. Small grains, on the other hand, trace the gas features because they are mostly coupled to the gas.

They further study that such asymmetry possibly leads to azimuthal variations in dust opacity at the millimeter wavelength because large dust grains are trapped at pressure maxima due to the local gas density enhancement. They also argue that the variation of dust temperature in the azimuthal direction might be responsible for generating azimuthal asymmetry. The incident angle of the stellar radiation to the disk is not uniform if LkH$\alpha$ 330 disk is warped such as HD 142527 disk \citep{casassus15a,marino15}, the disk is differently flared in the azimuthal direction, or the shadowing variation behind the asymmetric outer gap wall in the inner disk, causing azimuthal variation in the dust temperature. Therefore, the $H$-band emission depends on internal disk structure, while sub-millimeter is not. All of the possibilities mentioned above can explain the origin of the anti-correlation observed between $H$-band and 0.87 mm continuum emissions.  

The key questions whether or not a massive planet can form the spirals in scattered light and the sub-millimeter asymmetry will arise. \citet{bae16} simultaneously reproduced the observed spiral arms in NIR and asymmetry in sub-millimeter continuum by a single planet-mass companion at the outer disk region in the case of SAO 206462 transitional disk, and thus their works support in explaining the LkHa330 structure similar to SAO 206462. If this is the case, a massive planet orbiting at the outer disk could be a more plausible explanation for current architecture seen in the LkHa330. They also predicted that brightest peak and the spiral arms will show observable displacement over the next few years. The follow-up observations in NIR and sub-millimeter continuum are demanded for revealing the origin of the structure.
Details of the arm structures will be the subject of later studies in this series. Here, we simply stress that LkH$\alpha$ 330 has a spiral arm structure, a supportive evidence for the presence of a massive planet. In the next section, we discuss the possibility of grain growth due to a planet-induced pressure bump. 

\subsection{Implications of grain growth due to pressure bump} 
\label{growth}
Grain growth can be found observationally from the slope of the SED, expressed as $F_\nu \propto \nu^\alpha$ where $\alpha = \beta + 2$, at mm/sub-mm wavelengths. We thus use this law to estimate the dust opacity in the LkH$\alpha$ 330 disk. $\beta$ is related to composition, shape, and size distribution of the dust that accounts for thermal emission at mm/sub-mm wavelengths. Generally, $\beta$ lies between 0 and 1 for disks around low-mass T Tauri stars and almost every disk falls in the range of $\beta < 2$ \citep{beckwith91}. Note that the $\beta$ of the interstellar medium (ISM) is about 1.7 \citep{li01}. It is widely accepted that the explanation for $\beta_{\rm disk} < \beta_{\rm ISM}$ is primarily a grain growth effect in the disk, and other contributing factors such as the geometry of dust particles or an optically thick disk are less effective and don't alter $\beta$ to the same extent \citep{draine06}. \citet{draine06} also shows that the dust grows from $\mu$m to mm sized grains when $\beta$ decreases below unity. 

Here we estimate $\beta$ with our 0.87 mm continuum observation and a previous result from a 1.3 mm continuum image taken by CARMA \citep{isella13}. The integrated intensity in the $\lambda = 0.87$ mm and $1.3$ mm continuum observations in west side of the disk is 101 $\pm$ 14.1 and 35 $\pm$ 1 mJy, respectively, but 87 $\pm$ 12.7 and 18 $\pm$ 1 mJy in the east side of the disk, respectively. As a result, the estimated $\beta$ becomes 0.7 $^{+0.5}_{-0.4}$ in the west side and 2.0 $\pm$ 0.5 in the east side, respectively. $\beta$ in the west side is below unity and indicates that grain growth probably takes place as suggested from the work of \citet{draine06}. It is theoretically inferred that the dust grains grow to around 500 $\mu$m or even larger, because such sizes are required for $\beta$ to be less than unity for a power-law distribution of grain sizes ($a$), $n(a) \propto a^{-p}$, where n(a) represents the abundance of grains with a particular size and $p$ is a power index of the distribution \citep{dalessio01,natta04a,natta04b}. $\beta$ in the east side, on the other hand, shows a typical number of $\sim$ 2, indicating that grain growth is not so active than the west side. Note that \citet{dalessio01} also show that $\beta$ will become approximately 1.7 when the maximum size of dust grains is less than 30 $\mu$m. 

A pressure bump is considered as one possible explanation to promote grain growth since it locally traps dust material in the azimuthal direction and as a result accelerates dust particles growth \citep{barge95,birnstiel13}. If this is the case, the larger grains subsequently stay near the pressure bump while the smaller grains get uniformly distributed in the entire disk as seen in Oph IRS 48 \citep{marel13,marel15}. We note that grains are likely growing everywhere in the disk, but transport is concentrating large grains at the pressure bump, resulting in asymmetric structures. LkH$\alpha$ 330 has at least one spiral arm feature and an asymmetric structure about east and west side of the disk. Although there are several explanations for these complex structures, a large unseen planet naturally explains simultaneously both of the structures. Observations at longer wavelengths available in the future with ALMA or VLA will be greatly helpful for further investigations of grain growth and the connection between large sized grains and rocky cores of planets.   

\section{Summary}
\label{summary}
We have conducted $H$-band (1.6 $\mu$m) linear polarimetric observations and 0.87 mm continuum observations toward a transitional disk around the intermediate-mass pre-main sequence star LkH$\alpha$ 330. The observations show that a spiral arm potentially exists in the disk interior where planets form. Although several possibilities for the origin of the spiral arm have been proposed, a massive unseen planet would be the most plausible candidate according to previous simulations \citep{zhu12,isella13,dong15}.

The possibility of dust trapping and grain growth in the asymmetric disk component was investigated using the opacity index ($\beta$) through the observed SED slope between 0.87 mm and 1.3 mm continuum emission. As a result, a low opacity index, $\beta = 0.7^{+0.5}_{-0.4}$  is obtained in the west side of the disk, while the east side shows a typical value of $\beta = 2.0\pm0.5$ usually seen in the normal primordial disks, suggesting that grain growth probably takes place everywhere and only millimeter-sized grains selectively accumulates at the pressure maxima in the west side of the disk. Combining the results of near-infrared and submillimeter observations, we conjecture that the spiral arms were in the upper surface and an asymmetric (potentially arm-like) structure was in the mid-plane of the disk. Continuum observations at longer wavelengths such as centimeters may show how large dust grains connect to the formation of rocky cores of planets and near-infrared observations with so-called extreme AO are required for further exploring the dust distribution in the disk. 

\acknowledgments
M.T. is supported by Ministry of Science and Technology (MoST) of Taiwan (grant No. 103-2112-M-001-029). Y.H is currently supported by JPL/Caltech. This work is supported by MEXT KAKENHI No. 23103004.

\section{Appendix \\
$H$-band radial polarization} 
\label{appendix}
\begin{figure}[h]
\begin{center}
\includegraphics[scale=0.6]{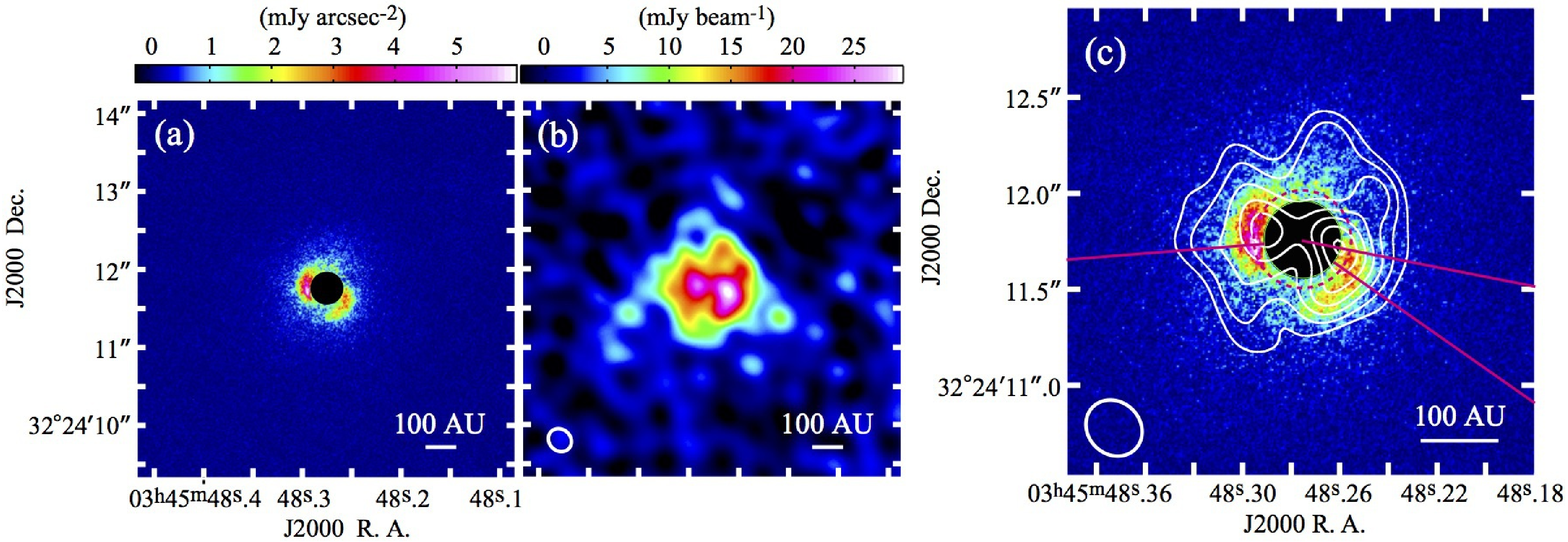}
\caption{Panel (a) shows the polarized intensity image in $H$-band by the Subaru telescope. A software mask with $r$ = 0$\farcs$2 is shown at the center. Panel (b) shows the continuum image at $\lambda$ = 0.87 mm taken by the SMA. The size of synthesized beam is $0\farcs31 \times 0\farcs27$, provided at the bottom left corner. Panel (c) represents a superposed SMA image on the $H$-band polarized intensity image in a larger scale. The color scale is the same as the one of Panel (a). The first contour is 8$\sigma$ and increases up to 20$\sigma$ with 2$\sigma$ steps. The red solid lines represent the directions of the three radial profiles provided in Figure \ref{fig:fig3}a and the red dashed-circle with $0\farcs25$ in radius passes on the emission peaks of the $H$-band and submillimeter continuum for the azimuthal profile as shown in Figure \ref{fig:fig3}b. North is up and east is to the left in all of the panels.}
\label{fig:fig1}
\end{center}
\end{figure}

\begin{figure}[h]
\begin{center}
\includegraphics[scale=0.5]{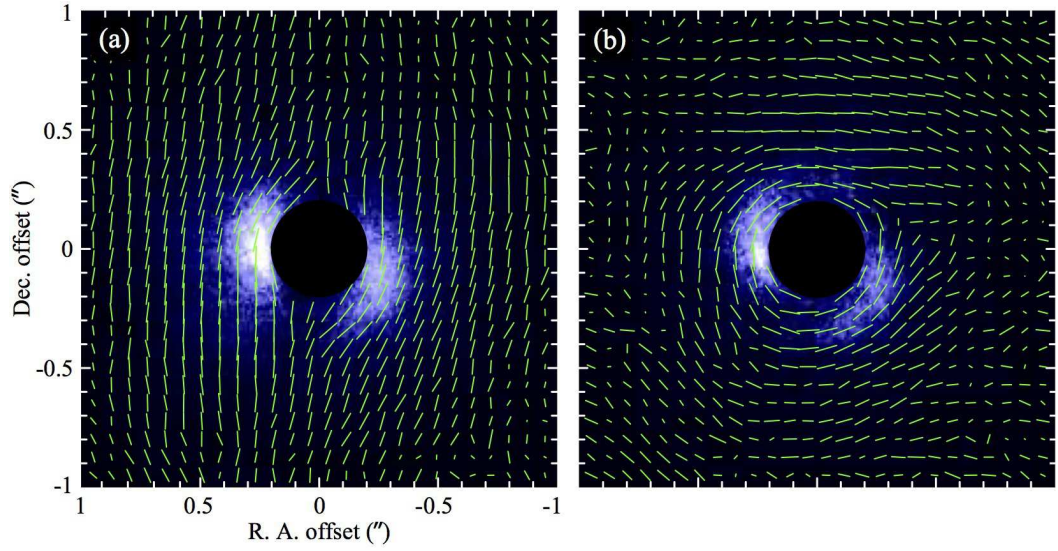}
\caption{$H$-band polarization vectors of LkH$\alpha$ 330 are superposed on the PI image with a software mask with $r$ = 0$\farcs$2. Panel (a) represents the polarization vector map before subtracting the polarized halo and panel (b) represents the same as panel (a) after subtracting the polarized halo. All plotted vectors' lengths are arbitrary for presentation purposes.}
\label{fig:fig2}
\end{center}
\end{figure}

\begin{figure}[h]
\begin{center}
\includegraphics[scale=0.65]{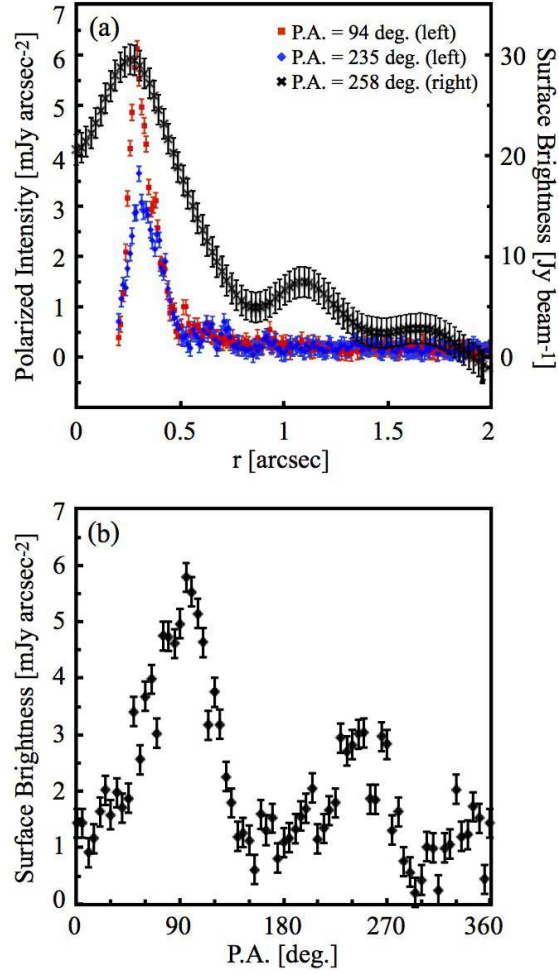}
\caption{Panel (a) represents the radial surface brightness profiles in the $H$-band observation along the directions of P. A. = 94$\arcdeg$ and 235$\arcdeg$ and submillimeter continuum observation along the directions of P. A. = 258$\arcdeg$ (see red solid lines in Figure \ref{fig:fig1}c). Note that only the region where $r > 0\farcs$2 is trustworthy in the $H$-band observation, while submillimeter data of the inner region where $r < 0\farcs$2 is provided in the plot. Panel (b) represents the azimuthal surface brightness profile at $r$ = 0$\farcs$25 ($\sim$ 60 AU) as indicated in the red dashed circle in Figure \ref{fig:fig1}c. The error bars show 1$\sigma$ in both the radial surface brightness  and the azimuthal surface brightness profiles.}
\label{fig:fig3}
\end{center}
\end{figure}

\begin{figure}[h]
\begin{center}
\includegraphics[scale=0.45]{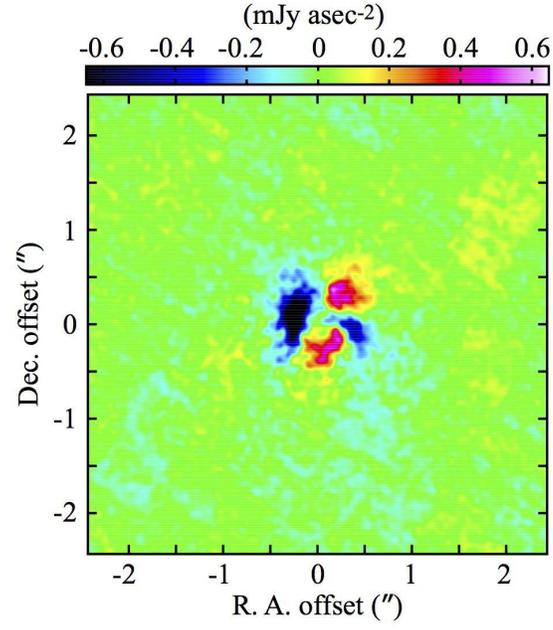}
\caption{Radial polarization (${\rm U}_\phi$) image at $H$-band whose polarization components are not orthogonal to stellar radiation. Note that the scale is approximately a factor of 10 smaller than the $H$-band polarization intensity shown in Figure \ref{fig:fig1}a, and thus the structures in Figure \ref{fig:fig4} are not affecting the discussed structures significantly.}
\label{fig:fig4}
\end{center}
\end{figure}

In addition to the $H$-band PI image provided in Figure \ref{fig:fig1}, we additionally report a radial polarization (${\rm U}_\phi$) map as a check for the polarization fraction that is not originated from the source (mainly from instruments). The ${\rm U}_\phi$ image corresponding to the PI image is provided in Figure \ref{fig:fig4}. In principle, polarization vectors have to be orthogonal to the incident radiation from the central star. Since the polarization fraction of ${\rm U}_\phi$ component is approximately 10 times smaller than the polarization intensity in the entire disk as shown in Figure \ref{fig:fig1}, the contribution of the instrumental polarization effect is negligible and not affecting the discussed structures significantly. Thus, we conclude that the observed polarized emission is originated from the disk of interest.

\clearpage

\clearpage

\clearpage

\clearpage

\clearpage
\end{document}